# Modelling Inter-species Molecular Crosstalks in Three Host-Parasite Systems by Expansion of their Sparse Information Space

**Paurush Praveen[1,2], Erfan Younesi [1,2], Martin Hofmann-Apitius [1,2§]**

[1]Bonn-Aachen International Center for Information Technology, University of Bonn, Bonn, Germany

[2]Department of Bioinformatics, Fraunhofer Institute for Algorithms and Scientific Computing, Sankt Augustin, Germany

[§]Corresponding author

Email addresses:

    PP: praveen@bit.uni-bonn.de

    EY: eyounesi@scai.fraunhofer.de

    MHA: mhofmann-apitius@scai.fraunhofer.de

# Abstract

**Background**

Parasitic diseases infect and kill many humans as well as domestic livestock around the globe, especially in the tropical countries. The emergence of cross-species interactions at the protein level is a part of molecular mechanisms that lead to such diseases. Comprehensive modelling can capture such interactions and could be useful to understand the pathophysiology of parasitic diseases and assist in identifying novel drug targets.

**Results**

Using a combination of databases, text mining, and predictive methods, we expanded the sparse space of protein-protein interactions in three parasitic diseases, namely malaria, sleeping sickness and cattle theileriosis. Analysis of the host-parasite interface in these three network models revealed significant similarities in molecular mechanisms underlying host's invasion, immunomodulation, and energy metabolism. The models not only contained the majority of well-known interactions, but also suggested new possible pathways in the inter-species protein interaction maps. Enrichment of these maps with drug-target information showed a plethora of druggable space to be explored, and led to the proposal of two new targets for each malaria and trypanosoma model.

**Conclusions**

It is demonstrated that, in the absence of sufficient experimental data, text-mining and predictive methods can successfully contribute to the expansion of the sparse biological information space, which in turn may lead to proposal of novel hypotheses and suggestions for future experiments.

# Background

Evolutionarily, parasites have the advantage to be locally adapted to the hosts system. Regardless of being generalist or specialist, parasites need to cross host species barriers [1]. Although it is well known that the host-parasite relationship is based on intimate interactions between the two species at the molecular level, the question is how many divergent and similar strategies at the molecular level have been developed by parasites to invade the host. As such, a holistic view of these interactions might reveal various aspects of the disease mechanism in the form of an interaction map. Under real world conditions, the host-parasite systems involve many cross-species molecular interactions as seen in case of bacterial and viral infections [2,3]; however, the focal point of most studies has been confined to specific singleton interactions due to the complexity of the multi-components involved. Although single-species interactome maps for parasitic organisms such as *Helicobacter pylori* [4], *Plasmodium falciparum* [5] and *Trypanosoma cruzi* [6] have been reported, less attention has been paid to the interactomes involving more than one proteome simultaneously. So far, major works in the field of host-parasite interaction belong to human-virus systems (e.g. a model for the human and herpes virus interactions at protein level [7]; development of a knowledgebase, namely VirHostNet containing virus and host interactions [8,9]; the HIV-1 human protein interaction database at the National Library of Medicine [10]) and to a limited extent to human-bacterial systems like Helicobacters or Staphylococcus. In an attempt to obtain the first landscape of human proteins interacting with viruses and bacterial pathogens, Dyer et al. (2008) used an integrative approach and constructed human-pathogen protein-protein interactions (PPI), generating the global view of infection strategies used by viruses and bacteria [11].

Unfortunately, similar interaction data do not yet exist for a wide range of parasitic infections; therefore, some alternative efforts are required to expand the information space for such interactomes. Protein interaction prediction, data- and text-mining techniques can be used to formulate a more sophisticated methodology for this purpose. These methods can achieve a higher degree of completeness and hence "real world relevance". For example, Rao and coworkers (2010) combined host and parasite PPI data, extracted by text mining of cerebral malaria-specific literature, with data from a number of predicted datasets and annotated the final interactome with Gene Ontology terms [12] .

Congregating data from different resources to expand the information space around these parasitic diseases is, however, a non-trivial task due to limited data availability. In the current work, we apply state-of-the-art text-mining technologies to biomedical literature [13,14] with the aim of expanding the interactome information space based on co-citation or other methods [15] together with PPI prediction approaches to explore the resultant space for the enrichment of disease mechanisms and pharmacological space around three protozoan diseases i.e. malaria, Human African Trypanosomiasis (HAT) and east coast fever. The purpose behind the selection of these three different systems was two-fold: to test the applicability of our approach to different combinations of host and parasite interactions, and to compare the early mechanisms of infection in these host-parasite systems.

Finally, we present new hypotheses on the pathology of the three systems and propose novel potential drug targets for malaria and for the HAT disease.

## Results and Discussion

We describe the host-parasite system as three-component machinery: the host interacting component, the parasite interacting component, and the host-parasite interacting complex creating an interspecies contact interface in the interaction map. To construct protein-protein interaction networks for such a system, it is necessary to obtain PPI data. Data on interspecies interaction are very sparse in existing databases. To compensate for the limited availability of information on protein-protein interactions between host and parasite, we used protein interaction prediction algorithms and text-mining methods to enrich the sparse PPI information space retrieved from protein interaction databases for all three parasites (Table 1). In fact, host-parasite complex (interface) interactions were available only for Plasmodium-Human from a published study [16].

Based on the data retrieved and prepared from databases, interaction prediction methods or text-mining, host-parasite protein interaction networks were created for all three parasitic systems (see Methods for details on prediction methodologies used). The network models generated by our workflow display the properties of biological networks in terms of clustering coefficient and small world property compared to random networks [17] (Table 2).

In the following, network models generated for specific host-parasite combinations are described and each network model is assessed by comparison to a set of known host and parasite interactions taken from the existing bibliome.

### *Plasmodium falciparum - Homo sapiens* Network Model

This model describes aspects of the pathophysiology of malaria. The hallmark of the malaria disease is the degradation of hemoglobin (iron-containing oxygen-transport metalloprotein in the erythrocytes). The host hemoglobin is degraded by the parasite in the trophozoite stage of its life cycle.

The data for the inter-species (Human- *Plasmodium falciparum*) interactions came from three different resources as described in Table 1. It should be noted that these interactions are disjoint and do not overlap.

*Validating the network model:* In order to check the conformity of the network model with the established knowledge, we extracted biological entities and pathways, shown to be involved in the parasitosis of malaria, from publications and compared them to our network. We were able to recover several known genes/proteins and pathways from the network model (Table 3; Additional file 1). Although this analysis was not exhaustive, the identification of a significant number of pathways in the network provides suggestive evidence that our network model comprises the majority of relevant interactions.

*Biological enrichment and inference of new knowledge:* Gene Ontology (GO) as a uniform knowledgebase is commonly used for gene function studies by means of GO term enrichment analysis. Identification of overrepresented GO terms in the network could help better understand the functional relevance of network elements or modules. GO analysis was performed for the host and the parasite networks as described in the Methods section (Additional file 2). The results of this analysis suggest that the functional spectrum can be confined to three general activities:

1. Host penetration and invasion: The host nodes at the network were found to have a fair overrepresentation of GO terms for vesicle and membrane-bound vesicle proteins (p-value: 3.5151E-4). During invasion, these host proteins may help the parasite to enter the host cell using the parasite apicomplex organelle [18,19]. Involvement of COP proteins in the network as well as enrichment for vesicle transport proteins (Protein piccolo, EEA1) implies dynamic cellular translocations. Interestingly, host proteins for inter-species interactions were also found to be on the interface. These proteins are involved in the processes by which one organism has an effect on another organism of a different species e.g. cell to cell adhesion or signalosomes. Protein kinases were also significantly enriched in the network (p-value: 2.1787 E-8 and 2.9590 E-8) indicating the existence of a substantial signalling component in the interaction network.

2. Host immunosurveillance escape: Presence of zinc homeostasis, TNF signalling and other cytokine pathways is evident in the network. Cytokines elicit inflammatory response and other disease-related activities [20]. In the course of infection by plasmodium parasites, the host cell is able to launch an immune response against the parasitic proteins via processes of recognition, response and modulation [21,22,23,24]. Figure 1a depicts network subgraphs showing the interactions between host and parasite proteins that lead to lymphocyte activation.

3. Taking over the host metabolic pathways: Purine salvage pathways were also detected to be significantly enriched in the network. The parasitic proteins interfere with the purine-related pathways of the host. The interference aims to meet the adenosine requirement of the parasite [25]. Purine salvage pathways in the intra-erythrocytic malaria parasite *Plasmodium falciparum* have been reviewed in terms of metabolism within the infected erythrocytes [26,27].

Moreover, there is an interference of the parasite with the proteins associated with spleen development. The spleen is the key site for removal of parasitized red blood cells together with generation of immunity and production of new red blood cells during malaria infection [28]. The ubiquitin system proteins detected in the model are probably involved in proteolysis by covalently binding to the target proteins [29]. It also contains interactions of parasite's proteins (PFC0495w, PF14 0075, PF08 0108, PF14 0281) with haemoglobin (Figure 1b). These proteins belong to the plasmepsin family (aspartic acid proteases) which can degrade the host hemoglobin and release toxic metabolites like hemozoin responsible for manifestation of the disease symptoms [30]. Interestingly, these interactions come from the prediction approach (prediction scores above 0.57) and their corresponding plasmepsin proteins show a high degree of sequence and protease domain similarities to human cathepsins.

In the parasite side, apicoplast proteins are related to the host cell invasion. Rhoptry proteins have roles in both biogenesis of rhoptry (a secretory organelle with enzymes for penetration process) and host cell invasion. They participate in the synthesis of fatty acids to enable the parasite to form the parasitophorous vacuole, imperative to a successful invasion of the host cell [31]. The metabolism-related proteins interfere with the host metabolic pathways such as energy metabolism and help the survival of the parasite inside the host cell.

In summary, it appears that the network model of plasmodium-human interactions represents the well-understood processes involved in host-parasite interaction with good coverage. Prediction of interactions among plasmepsin proteins demonstrates the applicability of the predicitve method for expansion of the information space by PPI prediction algorithms.

### *Trypanosoma brucei - Homo sapiens* Network Model

*Trypanosoma brucei* is responsible for the disease Human African Trypanosomiasis (HAT) which causes sleeping sickness by invading the central nervous system. The current model is an attempt to describe various aspects of this disease at the interface of host-parasite protein interactions. For the network construction task, we found no entries in public databases and thus it was imperative to obtain the modeling data from text mining and prediction methods. The interface between the host and the parasite consisted of 3368 interactions between 129 parasitic and ~2400 host proteins in the network (Table 1).

*Validating the network model:* The validity of the network was checked against the established knowledge by recovery of well-known nodes or pathways (Table 4). For example, Variant Surface Antigen (VSG) leads the immune evasion system [32,33,34]. The parasitosis is primarily associated with uncontrolled production of strong regulators like TNF, causing immune suppression, anemia, organ lesions and cachexia (weight loss, muscle atrophy, fatigue, weakness and anorexia). The VSG acts as a TNF inducing factor. The cysteine peptidase also plays an important role in the parasitosis by acting as immune depressant. The VSG also increases the levels of interleukins that help parasites to pass through the blood-brain barrier and invoke nerve tissue disorders and affect the circadian rhythm.

*Biological enrichment and inference of new knowledge:* Network annotation with the GO terms for the host and pathogen proteins provided a clue on involvement of interesting molecular crosstalks (Additional file 3). The interface network shows a large amount of parasite proteins interacting with the host pathways for cell cycle, apoptosis, signaling pathways and other important proteins like kinases. Protein-binding proteins and kinases formed the largest interface network component. The energy metabolism pathways were also found to be involved at the interface. The network analysis suggests that the parasite is able to control the host energy pathways for its own purposes. The immune system was well figured in the network, including the Variant Surface Glycoprotein (VSG antigen) that drives the immune evasion system [33,34]. The interaction between immunoglobin (IgM) and VSG was obtained only from text mining. Not really surprising, T-cell and B-cell activation [35] and regulation pathways were found to be influenced by the parasite. The variant surface antigen was found to interact with "CD5L_HUMAN", which is an IgM-associated peptide and may play a role in the regulation of the host immune system (Figure 2). Biologically, these interactions represent the parasitosis [36] and strong involvement of the immune mechanism at an abstract level.

Interaction of parasite proteins with human BCL-2 and BCL-3 implies that, for intracellular survival, the parasite must inhibit apoptosis of the infected cells. Interaction with host cell receptors such as TNF receptor 5 and chemokine receptor is apparently a successful strategy for penetrating the host cell machinery and immune suppression by the parasite. It seems that the parasite exploits TGF-beta in the early stages of infection to prepare the host cell for immune suppression [36,37]. Since the STAT family of transcription factors activates expression of immune system genes in human, interaction of trypanosoma proteins with human STAT proteins at the crosstalk interface might represent another aspect of well-concerted immunomodulatory actions taken by the parasite.

**_Theileria parva - Bos Taurus_ network model**
This model describes the 'East Coast Fever' disease which occurs in an animal host (i.e. cattle). The interaction network within the host (*Bos taurus*) was reconstructed in a fashion similar to the previous two models. During the analysis, attention was paid to the molecular crosstalk interface between the host and the parasite. The interface network was obtained through applying prediction methods and text-mining approaches as there were no database entries available for paired interactions between Theileria and cattle. The interface had 20 nodes from the parasite and 37 nodes from the host, forming 53 edges among them.

*Validating the network model:* The network model was validated by comparing its components to the established knowledge from literature, and recovery of the well-known proteins or signalling pathways from the network model (Table 5). For example, Ubiquitin-like protease expression has been observed during infection by *Theileria annulata* [38]. In T. parva infected T-cells, JNK and NF-κB are constitutively activated, further inducing activation of the transcription factors AP-1 and ATF-2. These, together with NF-κB, regulate the transcription of a number of genes that participate in the control of cellular proliferation and defence against apoptosis. The GTPases (among them Ras, Rac, Rho and Cdc42) are activated by different pathways, and also participate in the JNK activation pathway. The role of Src-relating kinases, casein kinase II and the possible interference of the parasite with many negative regulatory pathways can be observed in the model presented here. Two classes of MAPK, namely the JNKs (jun-NH2-terminal kinases) and the p38 family, mediate responses to cellular stress such as heat or osmotic shock, cellular injury, and inflammation [39]. Src family kinases and casein kinase 2 have also been known to play important roles in the pathology of the disease [40].

*Biological enrichment and inference of new knowledge:* Gene Ontology annotation of the host-parasite network revealed several categories of biological processes (Additional file 4) which are summarized as followings:

1. Penetration and cell trafficking: Functional enrichment of the interaction network at the host side shows that apart from metabolic activities, many proteins with receptor activity (P-value 7.67E-01) rank among the top 5 clusters. The receptor-mediated zippering of the target cell can provide the parasite with the opportunity to enter the host cell. Moreover, there are proteins in the network related to the membrane organization especially membrane invagination and endocytosis (e.g. RABGEF and NOSTRIN proteins) that are probably used in the process of cellular invasion by the parasite [41,42]. The presence of COP proteins in the network with significant annotation for transportation processes as well as Golgi-related complexes involvement support the model inferences. The PIM (Poly Immuno-dominant molecule) of Theileria which is part of the host-parasite interaction network may play a role in the regulation of cell adhesion.

2. Intracellular signalling: The role of kinases is highlighted in the model with protein kinase inhibitors and regulators (e.g. PKI, CSK and KAP). The abundance of intracellular signalling proteins (10 nodes) indicates the interference of the parasite in host intracellular processes.

3. Interference with the host cell cycle: The network was enriched for annotations of the cell cycle regulation and mitotic process-related functions with the presence of the proteins like JNK and RUBP. The TP01_0188 protein from Theileria was found to interact with 4 different host proteins (Figure 3). The TFDP, RBBP4 and BRMS1 are associated to cell division and apoptosis pathways whereas the CSE1L is responsible for intracellular transport of proteins, especially protein import into nucleus. These interactions indicate the interference of the parasite with the host apoptotic and cell cycle pathways.

The parasite layer displays the cell cycle control and regulation proteins including CDC2. This protein is involved in the process of cell division control but the network does not show any direct relation of parasite's CDC2 to any of the host proteins. This protein has high sequence similarity (63.45%) to its host counterpart, CDK2. Both proteins have very similar protein kinase domains. Based on this similarity, it may be speculated that they might have some common interactions. Since none of databases reports any direct interaction for CDK2 within the cattle as host, an analogy-based approach might help to infer that the parasite's CDC2 may have interactions similar to the host CDK2. The cell division protein kinase is involved in many pathways via phosphorylation/dephosphrylation steps. Curated interaction map of CDK2 in human indicates that this protein may be involved in many cell division and proliferation activities as demonstrated by the BioCarta pathways [43].

4. Evasion of immunosurveillance: The HSP (Heat Shock Protein) found in the network can probably refer to its involvement in the immune system response and evasion together with protection of the parasite during infection and inflammation [44,45]. Moreover, the nitric oxide synthase activity of NOSTRIN comes into the picture (P-value 1.4781E-2). The NOSTRIN pathway usually gets activated in the host's macrophages in response to the parasite infection [21]. Many of the parasite's metabolic proteins were also found to have interactions with similar proteins in the host including ATP and other energy related pathways.

*Characterization of bovine lymphosarcoma network model:* Cattles infected by Theileria parasite develop a disorder during a three- to four-week period that is similar to lymphoma (cancer of lymphocytes or lymphatic system) in humans. An interesting observation concerning the lymphosarcoma is that the parasite induces transformation of lymphocytes which is a reversible phenomenon in nature, i.e. lymphoma cells get back to their normal lymphocyte state when the parasite leaves the cell or dies out [39]. Therefore, it is intriguing to investigate which molecular mechanisms are used in the process of oncogenic transformation by the parasite and whether the reversibility of this process can serve as a model for better understanding of treatment strategies in human malignancies.

Here we characterize the network model of bovine lymphosarcoma and try to find biological processes similar to the human lymphoma. The model highlights the role of NF-kB under lymphosarcoma conditions in controlling the process of cell division and survival. It is well established that NF-kB plays a role in down-regulation of BCL-related genes in human [46]. An important observation in the model is the interference of Theileria proteins with the Akt pathway where Ras genes (via kinase cascades) are involved in the NFkB-IkB signalling, leading to taking control over the regulation of BCL-related genes. Thus, at this level, the model and the current knowledge in human lymphoma are in agreement and are both present a similar scenario (Figure 4). A possible explanation for this event provided by our network model is that the cell cycle or apoptosis pathways can be activated or deactivated based on the presence or absence of the corresponding inhibiting or activating enzymes which are directed by the parasite.

Accordingly, it can be hypothesized that when the parasite is inactive or dies, these activating or deactivating interactions disappear and hence alterations in the network signalling reverts, as there is no change at the gene level (i.e. permanent mutation) in the host cell. Based on the model, we can hypothesize that possible interference of the parasite signals with the Akt pathway and disruption of the NF kB - IF kB signaling may lead to the down-regulation of the oncogenic BCR-ABL fused gene. However, when the parasite is removed, probably the effects of this interference vanish and the hijacked pathways revert to their normal functioning status. Therefore, manipulation of suggested pathways in such a way that leads to restoration of the homeostasis in the intracellular signaling network may open up new opportunities for remediation of similar malignant disorders in human.

**Host-parasite networks: a comparative view**

Comparative study of host-parasite networks may reveal some interesting differences and similarities in the way that parasites choose to invade the host. Entry, multiplication, survival, and preservation of host function are of utmost importance to the success of parasites. Although a true comparative analysis involves alignment of networks, by only comparing the above-studied parasitic models at abstract level it can be seen that these parasites adopt a broad range of strategies to survive in the host's hostile environment.

The striking commonality among all the three parasitic models is the ability of parasites to pass through the barrier of host's immune surveillance by adopting different strategies such as immune system modification (Plasmodium), immune suppression (Trypanosoma), and oncogenic transformation (Theileria). Together with this, an activation of the nitric oxide pathway is observed as a common feature in all the three systems; the NO pathway is known to interact with immune regulatory mechanisms inside macrophages and other immune cells. Moreover, it has been experimentally shown that NO plays an essential role in host survival during *Trypanosoma cruzi* infection through suppression of the immune system [47]. Thus, the pivotal role of the NO pathway in diverse parasitic pathogenesis might indicate implications for therapeutical purposes.

Complementary to the NO pathway observation, the predominant interactions in the parasite-host network models involve survival processes. In all three models, significant metabolic interactions between parasite and host exists as parasites divert nutrients towards their own demands by taking control over host energy metabolism. Although a common underlying mechanism can be observed between the two apicomplexans (*Theileria parva* and *Plasmodium falciparum*), the mode of their parasitic action is different. Interactions involved in both cases affect many common pathways including the NF-kB signaling, cellular protein kinases and calmodulin pathways.

Taken together, the varied nature of the crosstalk that exists between these three host-parasite systems implies that molecular crosstalks between host and parasite are – not surprisingly - potentially the result of the coevolution of host-parasite genotype interactions [48], which increases the adaptation fitness and reduces the cross-species barrier.

**The drug space analysis and target discovery**

The complete drug-protein network, generated by merging drug information into the disease networks, provides an insight into the relations of the proteins with anti-parasitic drugs as well as the shared drug space across the inter-species and intra-species proteins. Some drugs were found to share the targets across the species boundaries (e.g. Aspirin, Figure 5). Apparently the drug space for malaria as well as trypanosomiasis is abundant and sparse since we did not find too many drug-protein interactions. In comparison, the average pharmacological density in the proteome of trypanosomiasis is pretty low (Figure 6).

In the available malaria parasite interactome about 89 proteins have been already targeted by the available drugs. For Plasmodium, the number of druggable proteins is estimated to be around 200 [49,50]. The remaining part of the parasite interactome is either undruggable or unexplored. However, the case is a bit different for the Trypanosoma interactome. Based on the drug-protein network, one can observe that the druggable proteome of Trypanosoma compared to Plasmodium has a larger share in its known interactome (~ 23%). From the viewpoint of druggability, both interactomes have ample sparse space to explore. This sparse area can be analyzed with model-driven approaches using rationales and features that define a druggable protein, leading to therapeutic application of network biology in discovery of new targets for these diseases. A large number of existing drugs were found to act on both Plasmodium and Trypanosoma parasites, speaking for the existence of a shared drug space between these models [51].

In order to prioritize possible drug target proteins in our host-parasite networks, primary rationales were set and the following features were used for prioritization: being parasitic-centric (network hubs), having low degree of similarity with host proteins, avoiding the proteins at the host-parasite interface, protein structure availability, and previous knowledge on druggability. Furthermore, these were evaluated by means of parameters such as sequence properties and clustering with existing drugs to find potential targets. Based on this method, it appears that superoxide dismutase, falcipain 2, and thioredoxin have good potential to serve as new drug targets in malaria, whereas in the case of trypanosomiasis, trans-sialidase and thioredoxin are proposed as possible target candidates. Therefore, the network approach not only allows for identification of target proteins similar to previously targeted ones, but also provides information on novel targets that are not linked to any known target.

## Conclusions

In conclusion, this work presents a first step towards elucidation of molecular models at the interface of parasite-host interaction using complementary methods, particularly when sufficient experimental data is not available. The network modelling approach has the advantage that the "known" or "understood" interactions and pathways can be expanded by new knowledge, crosstalks can be evaluated, and new regulatory components might be added and identified. Obviously, suggested models need to be improved as new data becomes available; however, the immediate outcome of network modeling approach is that it represents all the different facettes of the host-parasite interaction, going beyond the conventional "one parasite – one host" paradigm. Similar to comparative genomics in which data scarcity was compensated by orthology analysis, comparative network models might also provide a means for better completeness of the current information space.

## Methods

### Creating Information Space

The first phase focuses on creating an information space by pooling in data from different sources. At this point, the interaction data is retrieved using three different approaches; first, experimental data are retrieved from the existing protein interaction databases (BIND [52], DIP [53], HPRD [54], MINT [55] and IntAct [56]) via the interaction data integration system BIANA (Biologic Interactions and Network Analysis), an interaction database integration tool that integrates protein-protein interactions automatically across the above databases [57]. Second, extraction of gene/protein named entities from the literature was implemented on MEDLINE abstracts using a parasite-specific dictionary of protein names and their synonyms, which was compiled during the work. The text mining tool used for this purpose was ProMiner, a NER (Named Entity Recognition) tool [58], developed at the Fraunhofer Institute for Algorithms and Scientific Computing (SCAI). ProMiner uses a rule & dictionary based approach to recognize named entities in the text. The dictionary for cattle (*Bos taurus*) was available from a collaborative work between Fraunhofer SCAI and Department of Animal Science, University of Bonn, whereas the human dictionary was already available in the pre-compiled form. The ProMiner software was run over the MEDLINE abstracts with these dictionaries and results were indexed. To handle the two-entity class problem (detecting co-occurring proteins from two different organisms in an abstract) a Perl script was implemented to generate a list of abstracts which had the entities for two different organisms together with the entity name and the corresponding identifier. To avoid false positives, these abstracts were manually checked and a co-citation list was thus retrieved.

To grow the data further, a workflow was implemented to predict interaction between a pair of proteins across species based on three parameters: orthologue data, interacting domains, and gene ontology. The workflow starts with finding the orthologues of parasite proteins in the host. For this purpose, the In-Paranoid approach [59] was used. The orthologues of parasite proteins in the host organism were used to populate the initial interaction data. For each host orthologue, the neighbouring interaction partners within the host are detected from interaction databases, using them as seed nodes. Thus, a set of possible interaction partners for parasite proteins inside a host proteome is generated. This population of data is then enriched with the GO terms and PFAM domain IDs. The prediction program uses this data to compute overall interaction score.

The prediction feature set consists of the GO term similarity associated to the involved proteins A and B. The value is measured in terms of Jaccard Index (JI) as in Eq (1). Thus in ontology space, the value shows the closeness of proteins in terms of biological process, molecular function and cellular location. The values are normalized between 0 and 1:

$$JI(A,B) = \frac{|A \cap B|}{|A \cup B|}$$

(1)

A Jaccard Index of 0 means larger distance whereas 1 implies maximum closeness in ontology space.

Another feature is the interacting domain occurrence score (IDOS). The PFAM domains of the participant proteins can give a set of permuted pairs of domains. These pairs, when searched against a database of interacting domains, return occurrence or non-occurrence in the database and can contribute to the probability of their interaction. This occurrence score in Eq (2) can hence be used as a feature in terms of its log values for uniform distribution of all the features. The dataset for interacting domains was obtained from public databases viz. DOMINE [60]. The number of occurrence of a domain pair in the database is searched and scored (0 for 'found' and 1 for 'not found') and the number of interactions for a domain is counted (IN). These counts form the IDOS score. The score is normalized between 0 and 1 to avoid weight influence on the overall score.

The overall prediction score is a linear combination of these parameters. The constants are obtained from likelihood ratios for these features [61]. In this format of equation new features can be easily included to optimize the scoring function. Thus,

$$InteractionScore = (L_o \times O + L_g \times JI + L_d \times IDOS)$$

**(2)**

Where Lo, Lg and Ld represent the likelihood of interaction of proteins when the features of orthology, ontologies and interacting domains are supporting the interaction, where, O is the orthology score.

The likelihood computation is performed based on a gold standard set of true and false interactions, obtained from the study of Patil and Nakamura [61]. The likelihood ratio can be estimated using the Bayes rule as in Eq (3). All the parameter values are normalized between 0 and 1 to avoid biasness in overall score.

$$L = \frac{P(features | True)}{P(features | False)}$$

**(3)**

The high false positive rates, an inherent disadvantage of high throughput methods, can be overcome by the use of learning model. To filter the interaction data, GO terms and interacting domains were used as filtering features. WEKA platform was used for ranking the experimental interactions in order of their real world possibilities through learning-based classifiers [62]. A program to extract these features from the data set and write them in WEKA readable ARFF format was implemented. The data was then filtered using learning classifiers, namely Decision trees, Bayesian Networks, Random Forest, and SVM. Training the learning model was performed on 4916 true interactions and 4226 false interactions with 10-fold cross-validation followed by a test on 170 interactions (92 True and 78 False interactions). The interaction with majority votes from different classifiers was accepted as a true one.

All the generated data was then normalized and merged to generate the complete set of interactions in the system under study.

## Authors' contributions

PP designed the research, developed the computational method, analyzed data and drafted the manuscript. EY contributed in network and data analysis and drafted the manuscript. MHA conceived of the study, and participated in its design and coordination. This work was supported by B-IT (Bonn-Aachen International Center for Information Technology) foundation through funding of PP and EY.

# Acknowledgements

We acknowledge Dr. Vinod Kasam for his fruitful advice in developing rationales for drug target discovery and Theo Mevissen for his support in text-mining.

# Figures

**Figure 1 - Network models representing the crosstalk interactions between Plasmodium and human cells and their biological implications**

A. The sub-graph from the Plasmodium-Human network showing the lymphocyte activation interactions.

B. Four parasite proteins interacting with the host protein of hemoglobin alpha subunit. The parasite proteins belong to the protease family (red nodes represent the parasite proteins and green ones represent the host proteins).

**Figure 2 - Network models representing the immunomodulatory interactions of parasite within host's immune system**
  A. Subgraph showing various host proteins involved in different aspects of the immune system activation. Host proteins responsible for leukocyte activation, T and B cell activation, and provoking NK cells possibly interact with trypanosoma proteins. Red nodes represent parasite and green ones represent host proteins.
  B. Subgraph showing the involvement of parasite VSG with IgM-related peptide of the host. Red nodes represent the parasite proteins and green ones represent the host proteins.

**Figure 3 - Interaction subgraph in the east coast fever model.**

The model illustrates the attempt of the parasite in controlling over the host's cell survival processes.

**Figure 4 - Overall model for the interference of the *Theleria parva* in the *Bos taurus* interaction system**

**Figure 5 - The ligand interactions of some drugs for malaria**

The COX-1 of human interacts with Aspirin, and also with the parasite lactate dehydrogenase

**Figure 6 - Representation of druggable proteins in the proteome of parasites**
  A. plasmodium
  B. trypanosoma

**Figure 7 - High level work-flow diagram for the host-parasite interaction network modeling**

The workflow is segmented into the four phases; information space creation, filtering & normalization, modeling and model analysis.

# Tables

**Table 1 - Number of interspecies protein interactions retrieved using different methods for the three models**

| Data Source | Theileria-cattle | Plasmodium-human | Trypanosoma-human |
|---|---|---|---|
| Database | 0 | 455 | 0 |
| Text mining co-citation (post-curation) | 1 | 22 | 8 |
| Prediction | 47 | 833 | 3362 |

**Table 2 - Clustering coefficients for the network models and their corresponding randomized networks**

| System | Network | Randomized |
|---|---|---|
| Theileria-cattle | 0.426 | 0.073 |
| Plasmodium-human | 0.02 | 0.001 |
| Trypanosoma-human | 0.291 | 0.002 |

**Table 3 - Proteins and pathways attributed to the malaria disease in the available literature**

| PMID | Organism | Proteins/pathways | Status in the network |
|---|---|---|---|
| 19591795 | Parasite | PfAMA_1 | Present |
| 19641203 | Host | IFN, TNF, IL-2 | Present |
| 19380468 | Host | TNF, IL-6 | Present |
| 15137943 18532880 | Host and parasites | Protein kinase | Present |
| 17029647 | Host | Cytokine | Present |
| 15988316 | Host | Ubiquitin pathways | Present |

**Table 4 - Proteins and pathways attributed to the HAT disease in the available literatute**

| PMID | Organism | Nodes/pathways | Status in network |
|------|----------|----------------|-------------------|
| PMC260043 | Host | B cell and T cell activation | Present |
| PMC1364224 | Host | IFN gamma | Present |
| PMC2707606 | Host | NF-kB | Present |
| 17944830 | Parasite | VSG | Present |
| 19818332 | Host and parasite | Protein kinases | Present |
| 11237819 | Host | TNF | Present |

**Table 5 - Proteins and pathways attributed to the cattle's east coast fever in the available literature**

| PMID | Organism | Nodes/pathways | Status in network |
|------|----------|----------------|-------------------|
| 16441438 | Host | TashAT family | Present |
| 11207566 | Host and parasite | Protein kinases | Present |
| 10347800 | Host | PCD, TNF, NO pathways | Present |
| 15245750 | Host | IFK-B signalling pathway | Present |
| PMC2628565 | Parasite | PIM's role in cell adhesion | Present |

## Additional files

**Additional file 1** – The proportion of known as well as new pathways recovered from the network model of plasmodium-human

**Additional file 2** – GO analysis results for the host and parasite networks in the interaction model of plasmodium-human

**Additional file 3** – GO analysis results for the the host and parasite networks in the interaction model of trypanosoma-human

**Additional file 4** – GO analysis results for the the host and parasite networks in the interaction model of theileria-cattle

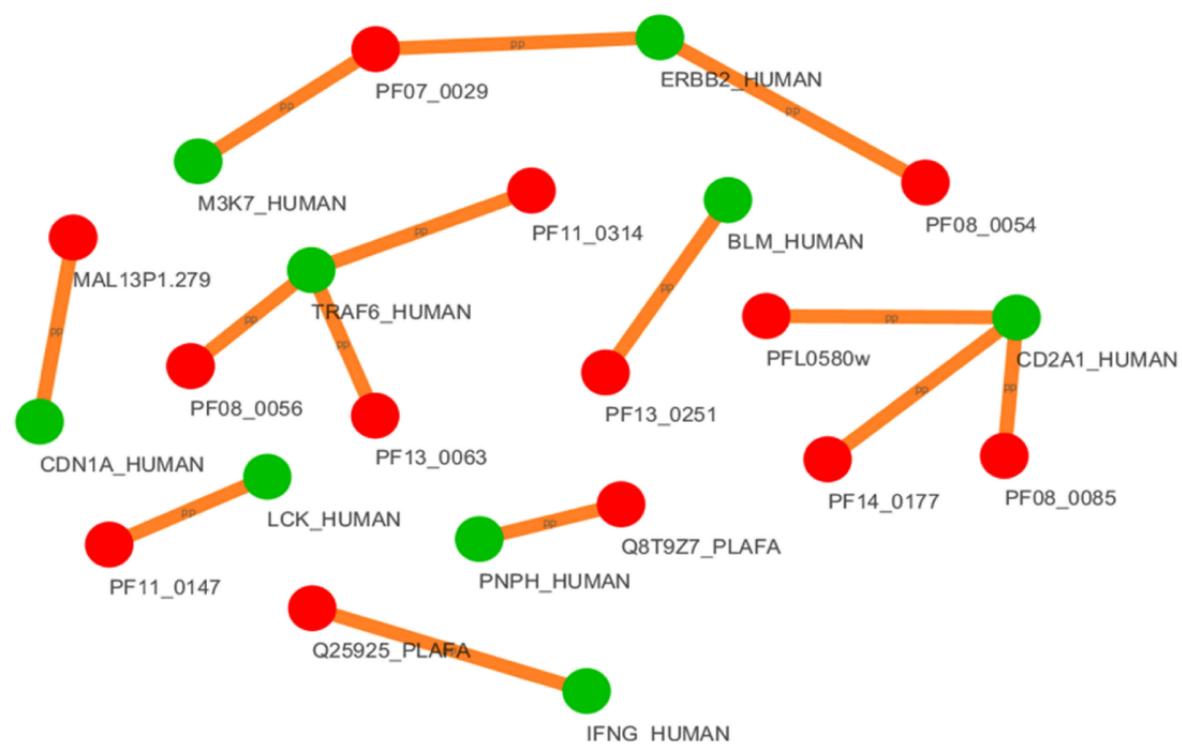 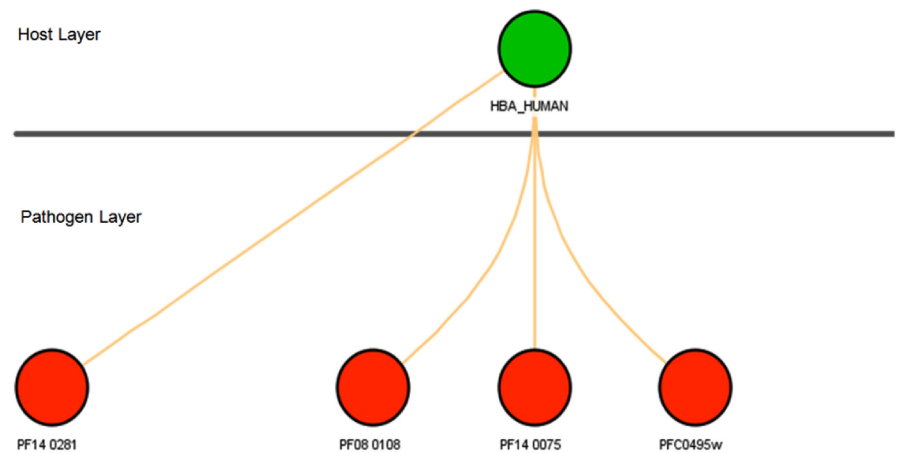

a

b

Figure 1

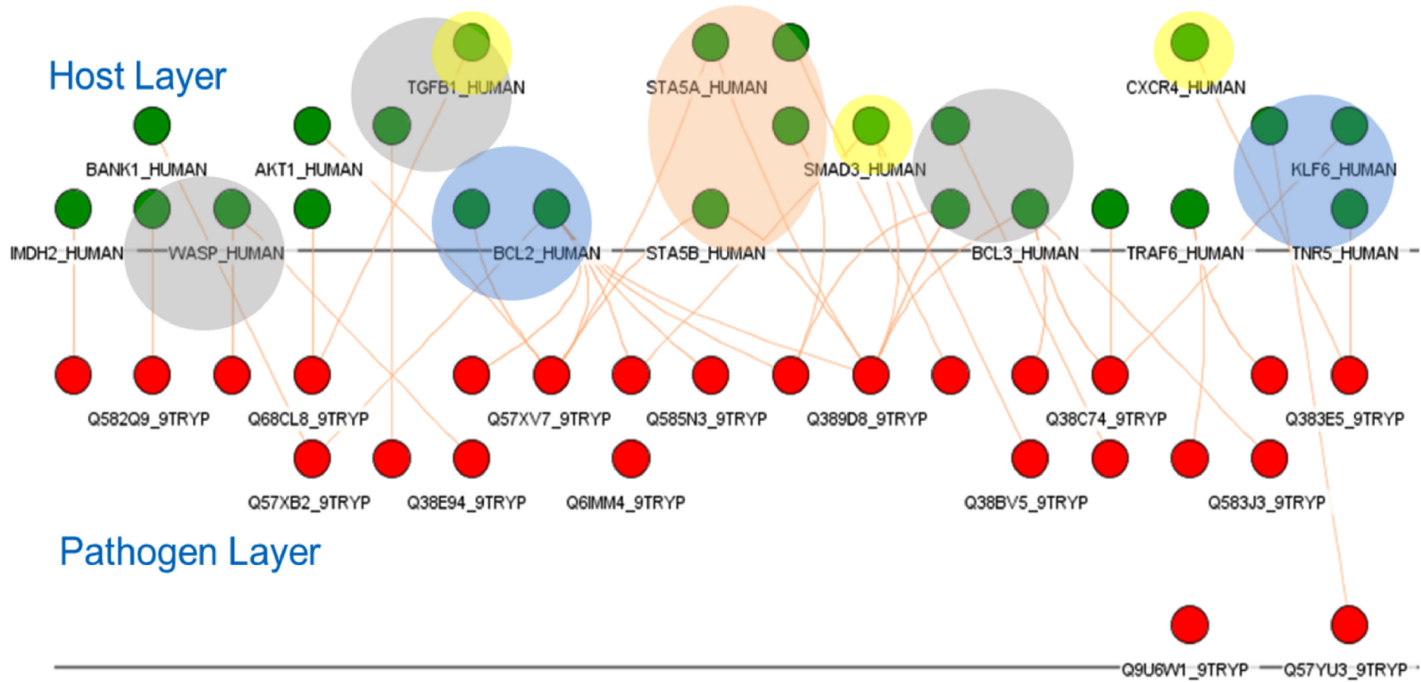
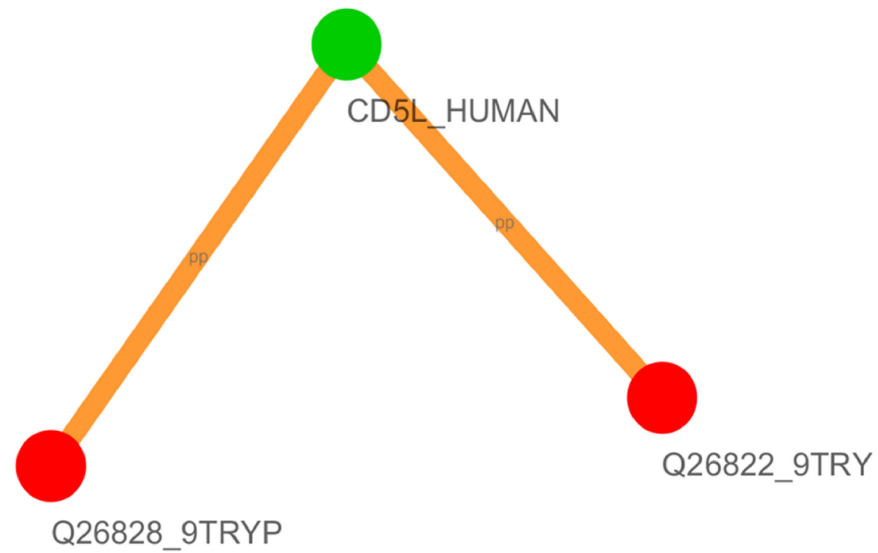

Figure 2

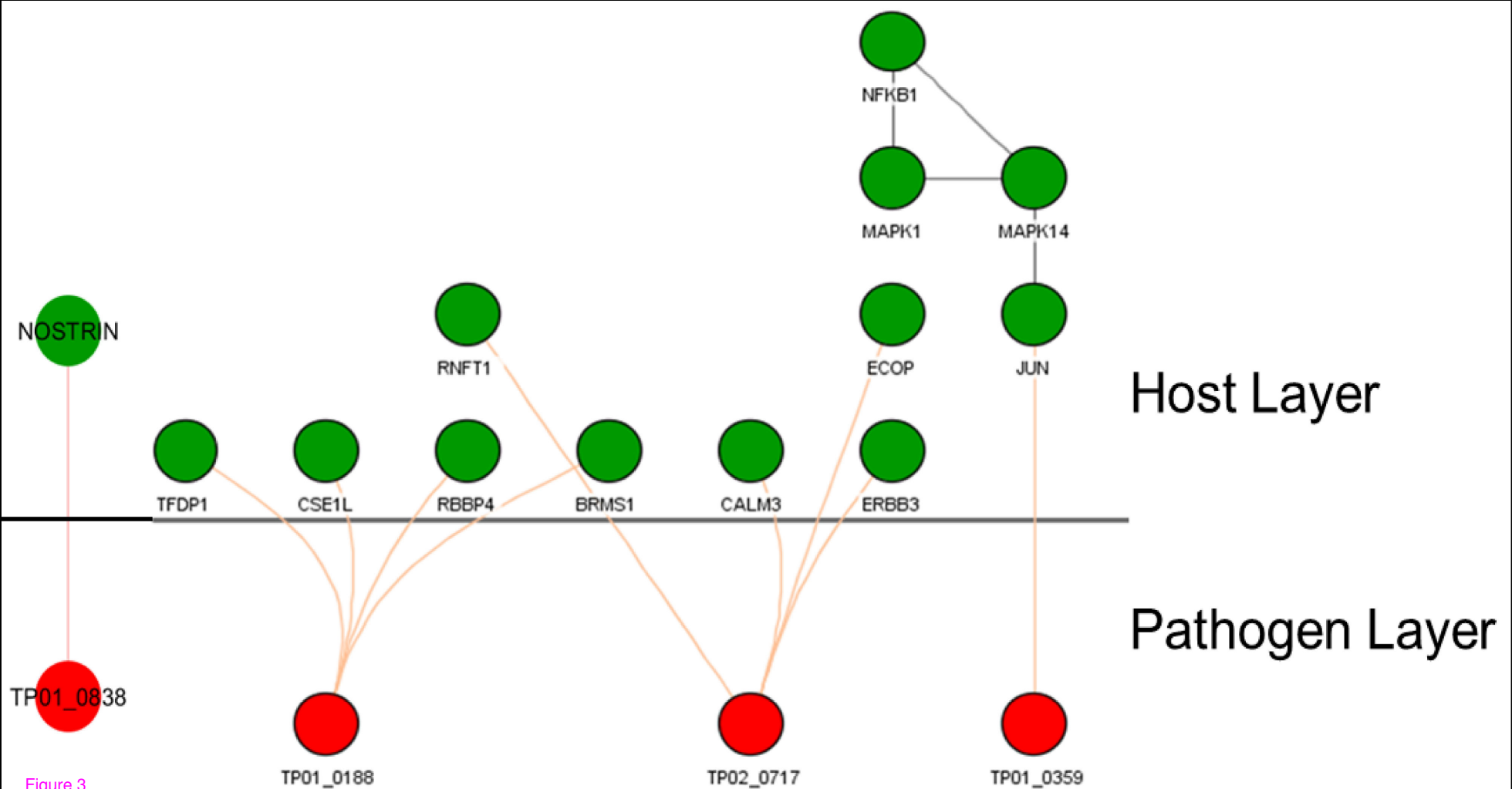
Figure 3

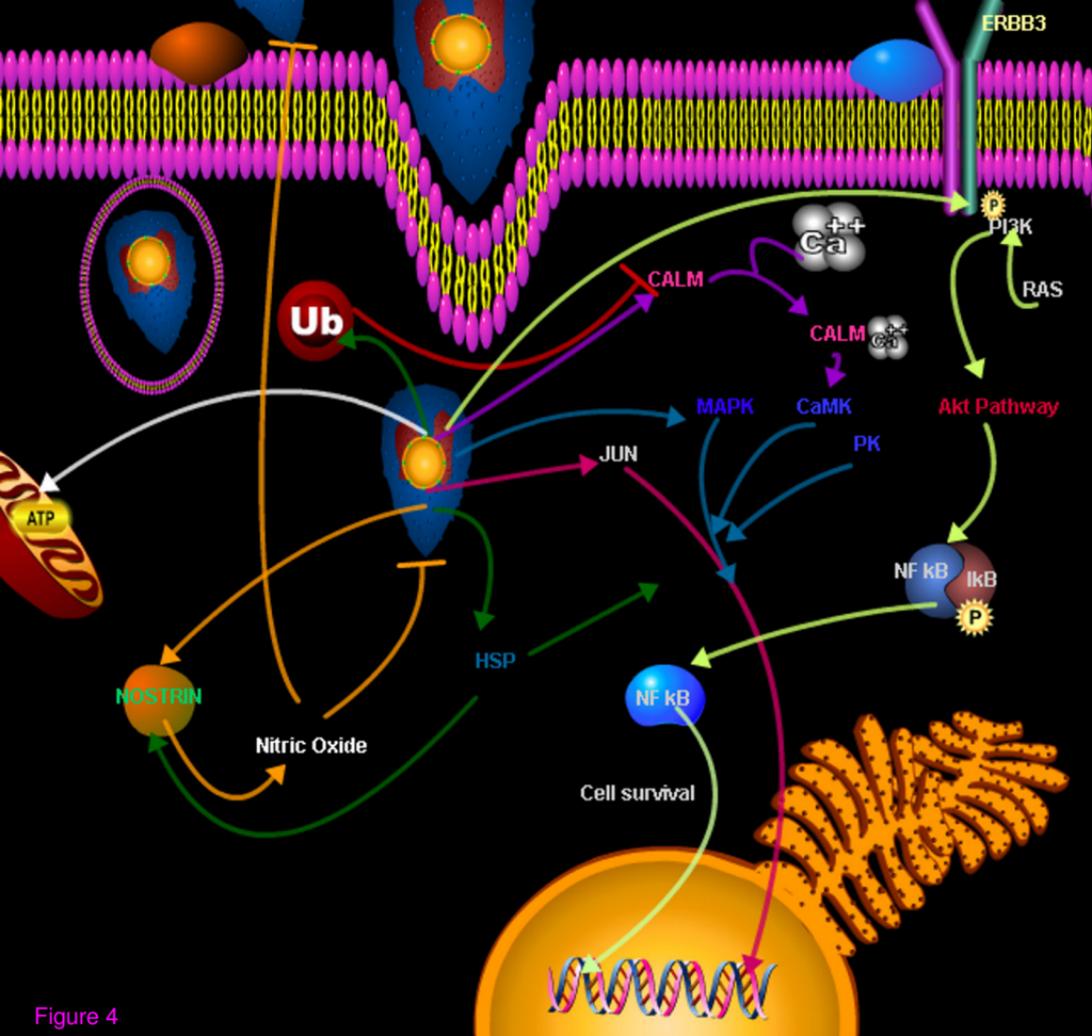

Figure 4

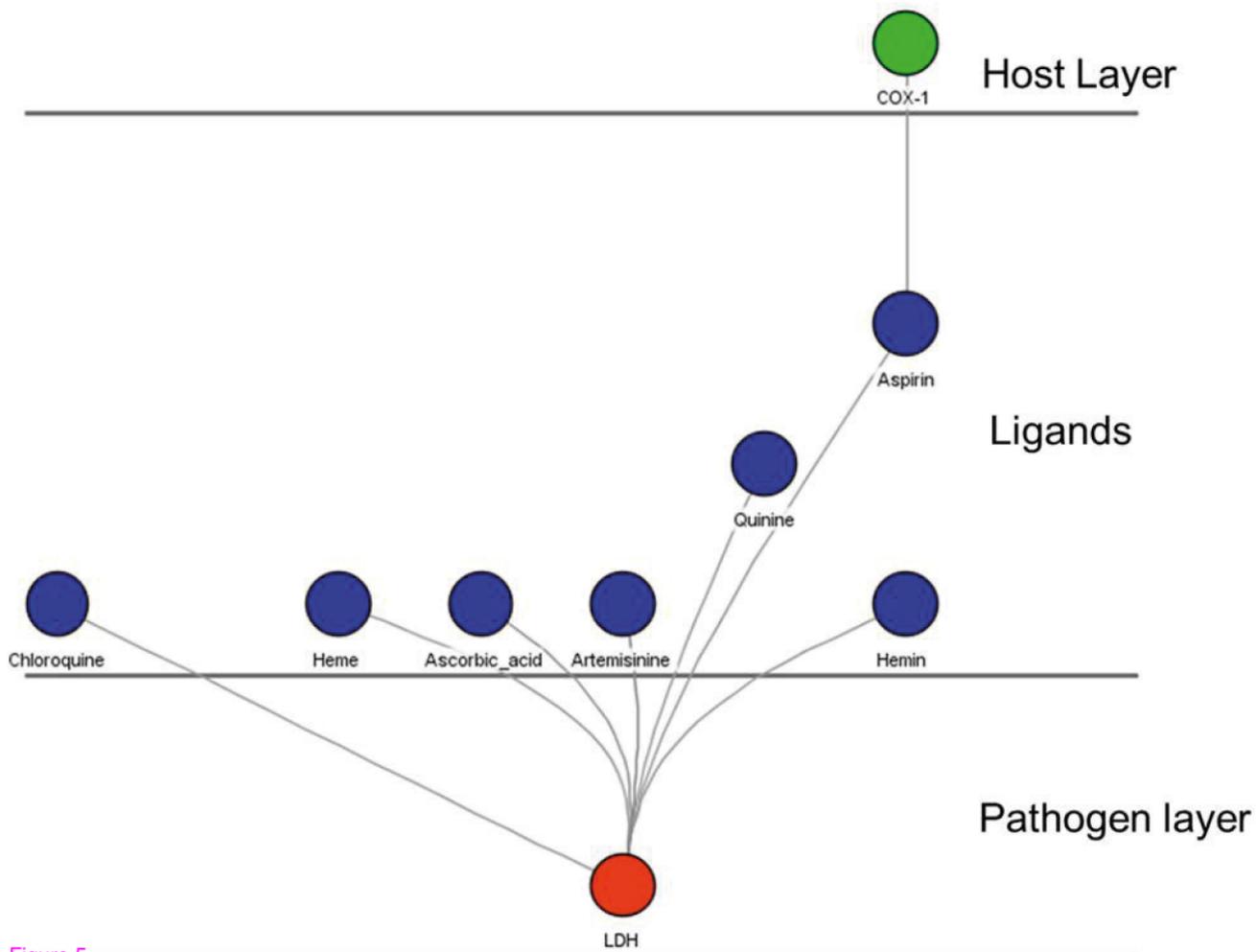

Figure 5

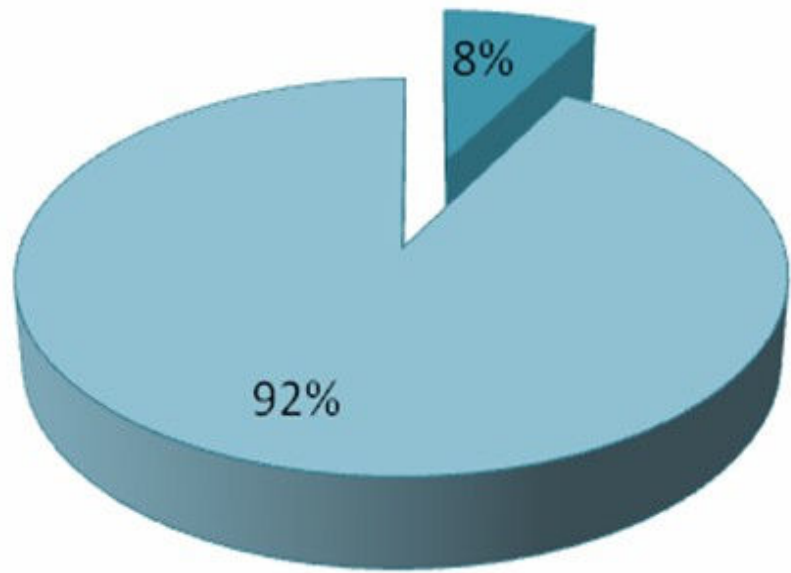 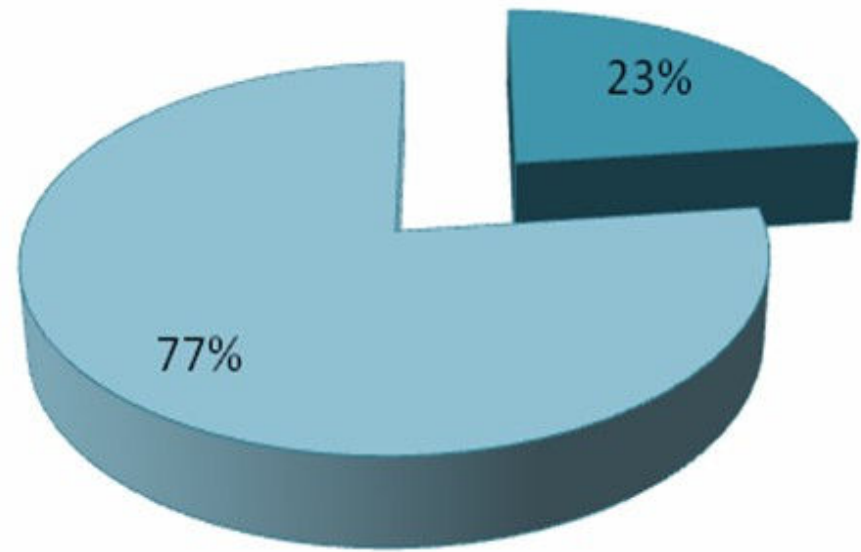

Figure 6

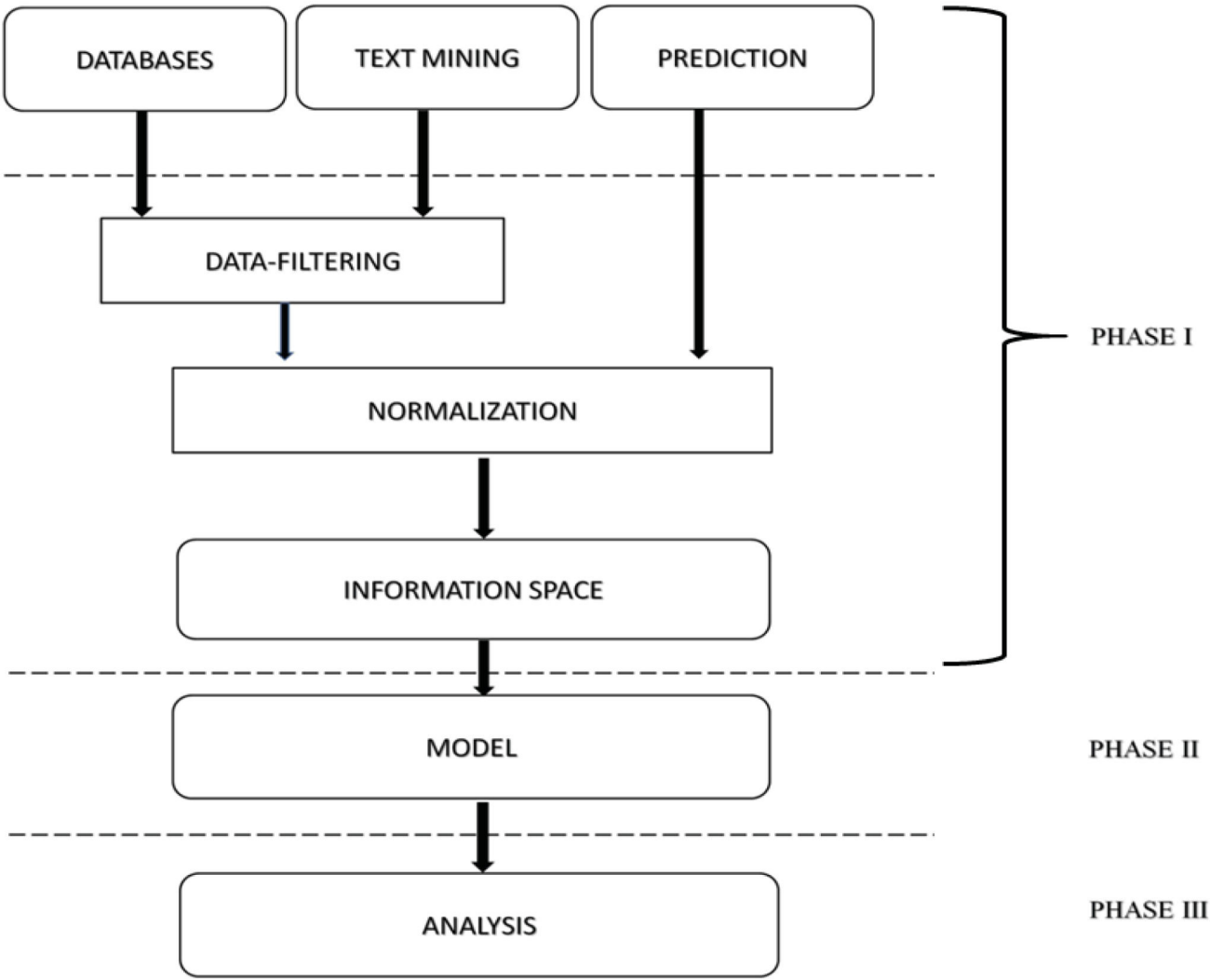
Figure 7